\documentstyle[12pt]{article}
\begin{document}
\begin{titlepage}
\title{A Broken Gauge Approach to Gravitational Mass and Charge}
\author{ T. Dereli\footnote{Leverhulme Visiting Professor.
Present address: Department of Physics, Ko\c{c} University, Istanbul, Turkey.} ,
R. W. Tucker \\  \\ {\small Department of Physics, Lancaster
University}\\ {\small Lancaster LA1 4YB, UK}}
\date{ 14 December 2001 }

\maketitle

\bigskip

\bigskip

\begin{abstract}
We argue that a spontaneous breakdown of local Weyl invariance
offers a mechanism in which gravitational interactions contribute
to the generation of particle masses and their electric charge.
The theory is formulated in terms of a spacetime geometry whose
natural connection has both dynamic torsion and non-metricity. Its
structure illuminates the role of dynamic scales  used to
determine measurable aspects of particle interactions and it
predicts an additional neutral vector boson with electroweak
properties.
\end{abstract}
\end{titlepage}

\def\e #1{\exp({#1})}
\def\lamtwo{{-c_2}}
\def\lamone{{c_1}}

\section{Introduction}

In this article we describe a symmetry breaking mechanism for the
emergence of gravitational mass and electric charge from
interactions that include gravitation. The view is taken that any
fundamental theory of masses should involve all the basic
interactions on some scale and that an effective theory of
gravitation in terms of a small number of relevant fields in
four-dimensional spacetime may be sufficient to discern a
mechanism of mass generation including gravitation. The theory
below will be formulated in terms of a spacetime geometry more
general than that adopted by Einstein, Weyl, Cartan and others
\cite{ 4a,4b,4c,4d} since it offers a new approach to the breaking
of a Weyl symmetry analogous to the breaking of internal
symmetries in the standard model of the fundamental particle
interactions. 

Since representations of the Weyl group are intimately connected with
the assignment of physical units to dimensioned quantities in any
theory with Weyl symmetry and the latter are related to
experimental predictions, it is useful to first make precise the
treatment of physical dimensions in a theory that maintains
spacetime diffeomorphism covariance throughout. Dimensional
analysis touches on the fundamentals of physical theory \cite{16c,
16d} and is particularly acute in theories such as gravitation
formulated in terms of some spacetime geometry. Tensor operations
(such as parallel transport, contraction, integration and
differentiation)   permit the construction of tensor equations for
any classical theory. The interpretation of the theory is in large
measure based on the correspondence of scalars with the results of
experimental operations. It is also a result of experience that a
hierarchy of broken symmetries of various kinds can be used to
classify the most relevant scalars in Nature.

Real spacetime geometry offers a framework to formulate particle
and gravitational field interactions on a manifold \cite{book}. By real
geometry, we mean here the assignment of a metric tensor field $g$
and a (Koszul) linear connection $\nabla$ on spacetime. The former
has Lorentzian signature while the latter is permitted to have
non-zero torsion

\begin{equation}
 T(X,Y)=\nabla_X\, Y - \nabla_Y\, X -[X,Y] \label{tor}
 \end{equation}
for any vector fields $X,Y$  and non-metricity  $S=\nabla g$.
 Given an arbitrary local basis of vector fields
$\{X_a\}$,  the most general
 linear  connection is specified locally
by a set of $16$ 1-forms  $\Lambda^a{}_b$ where:
\begin{equation}
\nabla_{{X_a}} \,X_b=\Lambda^c{}_b (X_a)\, X_c .
\end{equation}
Such a  connection can be fixed by specifying on spacetime the (2,0)
 symmetric metric tensor field
$g$,  the (2-antisymmetric,1) tensor field  $T$ and the (3,0)
tensor field $S$, symmetric in its last two arguments.  It is
straightforward to determine the connection in terms of these
tensors. Indeed since $\nabla$ is defined to commute with
contractions and reduce to differentiation on scalars, it follows
from the relation
\begin{equation}
X(g(Y,Z))=S(X,Y,Z)+g(\nabla_{X}Y,Z)+g(Y,\nabla_{X}Z)
\end{equation}
that
\begin{eqnarray}
2\,g(Z,\nabla_{X}Y)&=&X(g(Y,Z))+Y(g(Z,X))-Z(g(X,Y)) \nonumber\\
&&{}-g(X,[Y,Z])-g(Y,[X,Z])-g(Z,[Y,X]) \nonumber\\
&&{}-g(X,T(Y,Z))-g(Y,T(X,Z))-g(Z,T(Y,X)) \nonumber\\
&&{}-S(X,Y,Z)-S(Y,Z,X)+S(Z,X,Y)
\end{eqnarray}
where $X,Y,Z$ are any  vector fields. The  curvature operator
${\bf R}_{X,Y}$  of $\nabla$ defined   by
\begin{equation}
{\bf R}_{X,Y}Z=\nabla_
{X}\nabla_{Y}Z-\nabla_{Y}\nabla_{X}Z-\nabla{_{[X,Y]}}Z
\end{equation}
is  a type-preserving tensor derivation on the algebra of tensor
fields. The general (3,1) curvature tensor ${ R}$ of $\nabla$ is
defined by
\begin{equation}
{ R}(X,Y,Z,\beta)=\beta({\bf R}_{{X}{Y}} Z)
\end{equation}
where $\beta$ is an arbitrary 1-form. This tensor gives rise to a
set of local curvature 2-forms $R^a{}_b$:
\begin{equation}
R^a{}_b(X,Y)=\frac{1}{2}\,{ R}(X,Y,X_b,e^a)
\end{equation}
where $\{e^c\}$ is any local basis of 1-forms dual to $\{X_c\}$.
In terms of the contraction operator $\iota_X$ with respect to $X$
one has $\iota_{X_b}\,e^a\equiv
\iota_b\,e^a=e^a(X_b)=\delta^a{}_b$. In terms of the connection
forms
\begin{equation}
R^{a}{}_{b}=d\, \Lambda^a{}_b+\Lambda^a{}_c\wedge \Lambda^c{}_b .
\end{equation}
The general Ricci tensor is  $Ric\equiv R(-,-,X_a,e^a)$ with
curvature scalar  $$ {\cal R }\equiv G(e^a,e^b) \, Ric(X_b,X_a)$$
where $G(e^a,e^b)\,g(X_b,X_c)=\delta^a_c.$
In a similar manner the torsion tensor gives rise to a set of
local torsion 2-forms $T^a$:
\begin{equation}
T^a(X,Y)\equiv \frac{1}{2}\, e^a(T(X,Y))
\end{equation}
which can be expressed in terms of the connection forms as
\begin{equation}
T^a=d\, e^a+\Lambda^a{}_b\wedge e^b.
\end{equation}

Measurable quantities ${\cal O}[g,\nabla,C,\Phi,\ldots]$ are generally
functionals of the geometry, spacetime chains $C$ (such as
worldlines, worldtubes, spacelike hypersurfaces etc.), tensor and
spinor fields $\Phi$ carrying representations of local Lie groups
and constant physical parameters. For example local $p$-forms
$\alpha$ generate ${\cal O}[\alpha,C_p]=\int_{C_p}\alpha$ for some
$p$-chain $C_p$.
\def\dim#1#2#3#4{[#1_{#2}\,,#1_{#3}\,,#1_{#4}\,,\ldots  ]  }

\def\dimm#1#2#3#4{[#1_{#2}\,,#1_{#3},\,\ldots,#1_{#4}  ]  }

\def\tensor #1#2#3{#1^{#2}_{#3}}

\def\Tau{{\cal T}}
Let $\dim \lambda 1 2 3$ be a list of symbols that can be used to
define the  physical attributes of a quantity. This list encodes
the  {\it dimension} of the quantity. If $\dim d 1 2 3$ is a list
of (in general fractional) numbers then the $(p,q)$-type tensor
field $T^{(p,q)}_{\dim d 1 2 3}$ can be ascribed the arbitrary
dimension
$\lambda^{d_1}_{1}\,\lambda^{d_2}_{2}\,\lambda^{d_3}_{3}\,\ldots$.
Once dimensions are assigned  the theory is dimensionally coherent
provided only quantities with the same dimension are added
together. Tensor fields of different dimension can be multiplied
together and:

$$ {\tensor T {(p_1,q_1)} {\dim d 1 2 3}} \otimes  \tensor T
{(p_2,q_2)} {\dim {d^{\prime}} 1 2 3} =  \tensor T
{(p_1+p_2,q_1+q_2)} {[d_1 + d^{\prime}_1,d_2 + d^{\prime}_2,d_3 +
d^{\prime}_3,\ldots ]} .$$ Scalar fields are special cases in
which $\otimes $ is replaced by ordinary scalar multiplication and
for a  non-zero scalar field one may multiply by its reciprocal.
Once a local frame of vector and/or covector fields is choosen for
some spacetime region, the components of tensors are scalars
obtained by contraction. Although the $(p,q)$-type of a tensor
changes under the action of $\nabla$ and exterior differentiation
$d$, its dimension does not. Abbreviating $[T^{(p,q)}_{\dim d 1 2
3} ]\equiv [T]=
[\lambda^{d_1}_{1}\,\lambda^{d_2}_{2}\,\lambda^{d_3}_{3}\,\ldots]$
for any tensor $T$ then
$$[T(X,Y,\ldots,\alpha,\beta,\ldots)]=[T]\,[X]\,
[Y]\,\ldots\,[\alpha]\,[\beta]\,\ldots$$ $$[\nabla_X T]=[X] [T] $$
$$\left[\int_C\alpha\right]=[\alpha].$$ Units are numerical
measures assigned to physical  dimensions. A local choice of units
can be made by choosing a units frame $\dim {\alpha^0}
{[1,0,0,\ldots]} {[0,1,0,\ldots]} {[0,0,1,\ldots]}$ where
$\alpha^0_{[0,0,\ldots ,1,0,0\ldots]}$ is a non-zero constant
scalar field with dimensions $\lambda_k$. Such a field will be
called a {\it constant fiducial  scalar}. A priori there is no
reason to demand that it be dynamically related to any element in
the theory. In a theory with a choice of MKS dimensions
$[\lambda_1\equiv Mass,\,\lambda_2\equiv Length,\, \lambda_3\equiv
Time]$ a local units frame is given by the scalar fields
$[kg,m,sec]$ whose constant values at each event are $[1,1,1]$.
Just as an everywhere  non-zero scalar field with dimensions can
be used to change the dimensions of any quantity by
multiplication, so a general non-zero dimensioned scalar field can
be similarly used to change a units frame.

At any event $p$ in spacetime the scalar field quantity $\tensor
{\Phi} {} {\dimm d 1 2 n}$ is said to have the numerical {\it
value} $$ \frac{ \tensor {\Phi} {} {\dimm d 1 2 n}\vert_p }  {
\left( \alpha^0_{[1,0,0,\ldots]} \right)^{d_1}  \,\ldots  \left(
\alpha^0_{[0,0,\ldots, 1]} \right)^{d_n} } $$
 in a units frame
constructed from constant fiducial scalars. Clearly a constant
scalar has a constant value in such a frame. In a units frame
constructed from non-constant fiducial scalars the {\it value} of
a quantity may vary from event to event. Thus a meaningful
comparison of quantities with the same dimensions is obtained by
taking their ratio.

Most experimental measurements are interpreted in some canonical
units frame. In relativistic physics one may base a series of
measurements on the elapsed time between events according to a
choice of clock. In this way spatial intervals and particle
accelerations can be constructed consistent with special
relativity and Newtonian gravitation. In Einstein gravitation the
physics arises naturally from a (pseudo-)Riemannian spacetime
geometry based on a Levi-Civita connection  associated with a
metric tensor $g$.  It is convenient to induce dimensions from the
assignment $[g]=L^2$ where $L$ is a constant fiducuial {\it
length} scalar. Taking $c$ to be a constant {\it speed} scalar,
let $T=L/c$ be a constant fiducial {\it time} scalar. The constant
scalars $\Lambda_0$ (which we call {\it action}) and $Q_e$ (which
we call {\it electric charge}) are useful in the development of a
theory with charged masses. An MKS units frame $[m,sec,Joule-sec,
Coulombs] $ with values $[1,1,\hbar=1.0545 \times
10^{-24},e_0=1.6021 \times 10^{-19}]$ is traditional for the
theory. Whether such fiducial frames exist more generally, either
locally or globally on spacetime is a question for experiment and
can only be answered within the framework of physical laws.

It is clear that until an assignment of units is made only
dimensionless quantities are meaningful in a physical theory. It
has been suggested that the values of certain constant fiducial
scalars (e.g $\hbar, e_0$) arise from a particular choice of a
spacetime metric solution to the equations for gravitation and
that in the presence of dynamical scalar fields new units frames
can be choosen in which some of the so called {\it constants of
nature} (such as the Newtonian gravitational coupling constant
$G$) have values that vary with cosmological time \cite{3a,3b,3c}.
Einstein's metric theory of gravity offers an unambiguous way to define time
intervals and if $c,e_0$ and $\hbar$ are  constant scalars in such
a description  a complete edifice can be constructed in terms of a
single constant fundamental length scalar. Masses
$\mu=\frac{\hbar}{c\lambda_0}$ are given naturally in terms of
their Compton wavelengths $\lambda_0$ and  $G$ appears to be
constant in this framework. If the theory is generalised to
include dynamical scalars and a more general spacetime geometry
then the role of the fiducial scalars is more diffuse. Since a
dynamic theory of spacetime geometry of necessity has implications
for its own experimental interpretation the latter may depend on
which scale or epoch of the universe the theory purports to
describe.

\def\stressT{{\bf T}}

We take the view that, in the current epoch of the universe, the
interpretation of the gravitational mass of {\it particles}
associated with fields in a generalised theory be based on its
reduction to equations that involve the Einstein tensor
$Ein=Ric-\frac{1}{2}\,{\cal R}\,g$ of the Levi-Civita
($T=0,\,S=0$) connection associated with the spacetime metric $g$
satisfying :
\begin{equation}
 Ein=\Tau
\end{equation}
 for some source tensor $\Tau.$ With
$[g]=L^2$, then $[Ein]=[1]$ and the equation $\Tau =8\pi G
\stressT$ identifies $\stressT$ as the Einsteinian mass-energy
stress tensor. For any spacetime observer $Z$ with $g(Z,Z)=-c^2$
the scalar $\rho= \,\stressT(Z,Z)
> 0$ is a physical mass density. If the above equation admits a
static solution with
\begin{equation}
g=-c^2\,\left( 1-\frac{2\varphi}{c^2}+\ldots\right) d\, t \otimes d\,t
 + \left(1+\frac{2\varphi}{c^2}+\dots \right)\,\hat{g}
\end{equation}
 in local coordinates $(t,x^i)$ and with
  $\hat{g}=\sum_{i=1}^{3} d\,x^i \otimes d\, x^i$,
   then in the non-relativistic
weak field limit, $\varphi$ may be identified with the Newtonian
potential satisfying
\begin{equation}
(\partial^2_{x_1} + \partial^2_{x_2} + \partial^2_{x_3})\varphi
=4\pi G\, \rho.
\end{equation}
  The Levi-Civita geodesic motion of a neutral
test particle then  reduces to the Newtonian equation of
motion:
\begin{equation}
\frac{d\,
X_i(t)}{d\,t^2}=-\partial_{x_i}\varphi(X_1(t),X_2(t),X_3(t)).
\end{equation}
 In a regular three dimensional  spacelike domain ${\cal B}$ the
gravitational mass $M_G$  associated with $\varphi$ is given as
\begin{equation}
M_G=\frac{1}{4\pi G}\int_{\partial {\cal
B}}\hat{\star}\,d\varphi
\end{equation}
in terms of the Hodge map of $\hat{g}$.

In a theory with Weyl scaling it is also important to be able to
identify electric charge. Classical electromagnetism is associated
with a closed 2-form ${\cal F}$ and a closed 3-form $j$ on
spacetime:
\begin{equation}\label{max1}
d\,{\cal F}=0
\end{equation}
\begin{equation}\label{dj}
d\, j=0.
\end{equation}
 In a (rationalised) MKS units frame,  ${\cal F}$ is assigned
dimensions $[Q_e/\epsilon_0]$ where the fiducial scalar
$\epsilon_0$ is assigned dimensions $[\frac{{Q_e}^2 T^2}{M L^3}]$.
The conserved current associated with the Maxwell 2-form ${\cal
F}$ is then defined by the field equation
\begin{equation}\label {max2}
  j=d\,*(\epsilon_0 {\cal F})
\end{equation}
so $[j]=[Q_e]$. The electric 1-form field $E$ and electric charge
density $\rho_e$ defined by the observer $Z$ are $E = \frac{1}{c} \iota_Z
{\cal F}$ and $\rho_e = \frac{1}{c} \iota_Z(*j)$, respectively.
Since ${\cal F}$ is closed there exists a potential 1-form ${\cal
A}$ on a regular domain of spacetime such that ${\cal F}=d\, {\cal
A}$. If there exists a static field configuration with $E=d\,
\varphi_e$ then the electrostatic potential $\varphi_e$ satisfies
\begin{equation}
(\partial^2_{x_1} + \partial^2_{x_2} + \partial^2_{x_3})\varphi_e =
\frac{ \rho_e}{\epsilon_0}
\end{equation}
 in the above metric in the weak
gravitational field limit. The electric charge $Q_e[\varphi_e,{\cal B
}]$ in Coulombs associated with $\varphi_e$ in a regular spacelike
three dimensional domain ${\cal B}$ is given as
\begin{equation}
Q_e[\varphi_e,{\cal B }]=\epsilon_0\int_{\partial {\cal
B}}\hat{\star}\,d\varphi_e
\end{equation}
 in terms of the Hodge map of $\hat{g}$.

The fundamental difference between {\it mass} and {\it electric
charge} arises from the structure of the sources $\Tau$ and $j$.
In the absence of local Weyl symmetry the structure of $j$ depends
on other fields $\Phi$ carrying representations of some $U(1)$
group. Thus if $\theta $ is an  arbitrary real scalar field on
spacetime and $e$ some {\it charge parameter} with
$[e\,\theta]=1$,
\begin{equation}
\Phi\mapsto \e {-ie\theta}\,\Phi
\end{equation}
 under the
$U(1)$  group action.
 If the above field
equations  (\ref{max1}),(\ref{max2})  are  generated from a
suitable locally $U(1)$ invariant action in which the 1-form
${\cal A}$ carries a representation of a $U(1)$ connection then
charge conservation is guaranteed and the field equations are also
$U(1)$ invariant. In a locally Weyl covariant theory it is more
natural to work with a dimensionless connection $A$ and recover
the Maxwell potential  ${\cal A}$, Maxwell field ${\cal F}$ and
MKS electric charge sources by transforming with a fiducial scalar
having dimensions $[Q_e/\epsilon_0]$.

\def\DD{{\cal D}}
\def\D{D}

To implement a theory of dynamic mass generation for the current
epoch we seek a mechanism to break a theory with local Weyl
symmetry.  To this end we must first construct a locally Weyl
covariant theory of gravitation where certain field elements
$\Phi$ also carry representations of the scaling group. Thus if
$\sigma$ is an arbitrary real scalar field on spacetime
\begin{equation}
\Phi\mapsto \e {-q\sigma}\,\Phi
\end{equation}
 under the Weyl group action.
The dimensionless real constant $q$ is called the {\it Weyl or scale
charge parameter} of the representation. Scale charges are
relative to the representation carried by a class of metric
tensors $[g]$, elements of which are equivalent under
\begin{equation}
g \mapsto \e {2\sigma}\,g.  \label{Weylg}
\end{equation}
We assign zero dimensions to $\sigma$ so that elements on the same
Weyl group orbit all have the same dimension. A Weyl connection
can be represented by a dimensionless 1-form $Q$ such that
\begin{equation}
Q \mapsto Q +d\,\sigma
\end{equation}
under a local Weyl transformation. In
terms of $Q$ the exterior Weyl covariant derivative of a generic
p-form $\Phi^p_{q}$  with Weyl charge $q$ is defined to be:
\begin{equation}
\DD\Phi^p_q=\D\Phi^p_q +q\,Q\,\wedge \Phi^p_q\label{DD}
\end{equation} where $\D$ is the exterior covariant derivative that
maintains the dimension, scale charge and  covariance of
$\DD\Phi^p_q$ under all local group transformations in the theory.
Thus under the Weyl group $\DD\Phi^p_q\mapsto \e
{-q\sigma}\,\DD\Phi^p_q$. For pure Weyl scalars $\D=d$.

A class of Hodge maps $[\star]$ is associated with $[g]$. One may
readily verify from (\ref{DD}) that

\begin{equation}\DD\star\Phi^p_q=\D\star\Phi^p_q +(q-(4-2p))\,Q\,\wedge
\star\,\Phi^p_q\label{starDD}.\end{equation}
If one denotes $\Phi^p_{q}$ by $\left\{ {}^p_q \right\}\Phi$ then
it is also straightforward to verify the following rules:

\def\pq #1#2{  \left\{ {}^{#1}_{#2} \right\}  }

$$ \star\left( \pq p q \Phi\right)=\pq {4-p}
{q-(4-2p)}\left(\star\,\Phi\right ) $$

$$ \DD\left( \pq p q\Phi\right) =\pq {p+1}
{q}\left(\DD\,\Phi\right ) $$

$$\left(\pq {p_1} {q_1}\,\Phi_1  \right)\wedge \left(\pq {p_2}
{q_2}\,\Phi_2  \right) = \pq {p_1+p_2} {q_1+q_2}\,\left(\Phi_1
\wedge \Phi_2 \right). $$

\def\RXY{{\bf R}_{XY}}
In order to construct a locally Weyl invariant action principle
for our models we  take the dynamic Weyl connection 1-form $Q$
proportional to the metric trace of the non-metricity tensor
$S=\nabla g$ in a geometry with:
\begin{equation}
S = 2\epsilon\,\, Q \otimes g\label{nonmeteqn}
\end{equation}
where $\epsilon^2=1$. In order to induce the appropriate behaviour
of $Q$ under a change of Weyl gauge, we adopt the Weyl
transformation rule
\begin{equation}
\nabla \mapsto \nabla  \label{weyldel}
\end{equation}
 for the connection.  From
(\ref{Weylg}) this induces the transformation:
\begin{equation}
Q\mapsto Q+\epsilon\, dr\,\sigma
\end{equation}
so we take $\epsilon=1$. Since $\nabla$
preserves Weyl covariance the torsion tensor given by (\ref{tor})
remains invariant as does the curvature operator
$\RXY$.{}\footnote{The more general rule $\nabla \mapsto \nabla+
d\,f \otimes$ acting on arbitrary tensor fields is possible and
also preserves $\RXY$  but serves only to complicate the Weyl
transformation properties of the torsion tensor \cite{2c}.}
In a field of local {\it  orthonormal} co-frames $\{e^a\}$ with
$g=\eta_{a\,b}e^a\otimes e^b$,  one has under a Weyl scaling
\begin{equation}
e^a\mapsto \e \sigma \,e^a.
\end{equation}
 Thus the torsion 1-forms $T^a$
transform as
\begin{equation}
T^a\mapsto\, \e \sigma \,T^a.
\end{equation}

\section{Weyl Invariant Actions}

To facilitate the derivation of the field equations from an action
it is convenient to adopt a class of local g-orthonormal
co-frames  $\{ e^a \}$ and their duals $\{ X_a \}$. In the absence
of internal gauge symmetries $D$ is then determined by the
 connection 1-forms $\{\Lambda^{a}_{\,\, b}\}$ in such bases.
Since $\eta_{a\,b}=g(X_a,X_b)$ and
$(D\eta_{a\,b})(X)=(\nabla_{X}g)(X_a,X_b)$ for all $X$:
\begin{equation}
D\eta_{ab} = d\eta_{ab} - \Lambda^{c}_{\,\, a}\eta_{cb} -
\Lambda^{c}_{\,\, b} \eta_{ac} = -(\Lambda_{ab} + \Lambda_{ba}) =
-2 Q_{ab},
\end{equation}
 in terms of the non-metricity 1-forms $Q_{ab}$.
Thus in a geometry with $\nabla_X g=2\,Q(X)\,g$ it follows that
\begin{equation}
Q_{ab}=-Q\,\eta_{ab}. \label{constraint}
\end{equation}
A geometry with this $S$ (but zero torsion)  was first suggested
by Weyl \cite{4a,4b} and provided the precursor to the modern gauge approach to
particle interactions. Under the local Weyl scalings  above:
\begin{equation}
 Q_{ab} \mapsto  Q_{ab}-\eta_{ab}\,
d\sigma.
\end{equation}
and it follows from the definition of the connection forms (in any
basis) that  $ [\Lambda^a{}_b]=1,\, [Q]=1,\, [ Q_{ab}]=1$. The
connection 1-forms in any  orthonormal basis can be decomposed
into their anti-symmetric $  \Omega_{ab} $ and symmetric parts $
Q_{ab} $,
\begin{equation}
 \Lambda_{ab} = \Omega_{ab} + Q_{ab}.
\end{equation}
The anti-symmetric part can be further decomposed in a unique way
according to
\begin{equation}
\Omega_{ab} = \omega_{ab} + K_{ab} + q_{ab}.
\end{equation}
Here the  (torsion-free) Levi-Civita connection 1-forms
$\omega^{a}_{\,\,b}$ satisfy the structure equations
\begin{equation}
de^a + \omega^{a}_{\,\,b} \wedge e^b = 0.
\end{equation}
The contortion 1-forms
$K^{a}_{\,\,b}$ fix the torsion:
\begin{equation}
K^{a}_{\,\,b} \wedge e^b =
T^a,
\end{equation}
 and
\begin{equation}
q_{ab} = -\iota_{X_a}Q_{bc} e^c + \iota_{X_b}Q_{ac} e^c.
\end{equation}
In the presence of
torsion and non-metricity the exterior covariant derivative $D$
above satisfies
\begin{eqnarray}
D*e_a &=& *(e_a \wedge e_b) \wedge T^b +4 Q \wedge *e_a, \nonumber \\
D*(e_a \wedge e_b) &=& *(e_a \wedge e_b \wedge e_c) \wedge T^c + 4 Q
\wedge *(e_a \wedge e_b), \nonumber \\
D*(e_a \wedge e_b \wedge e_c) &=&
*(e_a \wedge e_b \wedge e_c \wedge e_d) T^d + 4 Q \wedge *(e_a
\wedge e_b \wedge e_c), \nonumber \\
 D*(e_a \wedge e_b \wedge e_c \wedge
e_d) &=& 4 Q  *(e_a \wedge e_b \wedge e_c \wedge e_d)
\end{eqnarray}
 in terms of the Hodge map with $ e^0 \wedge e^1 \wedge e^2 \wedge e^3 = *1.$
These relations prove of value in the following sections.
\bigskip

\def\bfLam{{\bf \Lambda}}
\def\S{{\cal S}}
\def\c{\gamma}

An action $\S=\int_M\,\bfLam$ is locally Weyl invariant for some
4-form $\bfLam$ if, for some 3-form $\psi$ (with compact support
on spacetime $M$ ), the transformation $$\bfLam\mapsto \bfLam +
d\psi$$
 is induced by Weyl scalings.
We begin by considering a Weyl invariant action
$\S[g,\nabla,\alpha]$ constructed with the aid of a real dynamic
scalar field having $[\alpha]=L^{-1}$ and Weyl charge 1,
$\alpha\mapsto \e {-\sigma}\,\alpha $.  Since metric variations
are induced by orthonormal co-frame variations we write:
\begin{equation}\label{act1}
  \S[\{e^a\}, \{\Lambda^a{}_b\}, \alpha] = \int_M \bfLam_1
 \end{equation}
 with
  $$ \bfLam_1 = \bfLam_0\left( \alpha^2 R^{a}_{\,\,b} \wedge *(e_{a} \wedge
e^{b}) -\frac{\c}{2} {\cal{D}}\alpha \wedge *{\cal{D}}\alpha
-\frac{1}{2} dQ \wedge *dQ\right) $$
 The coupling $\c$ is a real
dimensionless constant while $\bfLam_0$ is a constant with the
dimensions of action.
Co-frame variations of (\ref{act1}) yield the gravitational field
equations
\begin{equation}  \alpha^2 G_a = \c \tau_{a}[\alpha] + \hat\tau_{a}[dQ]
\label{EIN1}
\end{equation}
where the generalised Einstein 3-forms
\begin{equation}
G_a \equiv
- R^{b}_{\,\,c} \wedge *(e_a \wedge e_b \wedge e^c)
\end{equation}
and the source 3-forms are given by
\begin{equation}
\tau_{a}[\alpha] = \frac{1}{2}(
\iota_{X_a}{\cal{D}}\alpha *{\cal{D}}\alpha + {\cal{D}}\alpha
\wedge \iota_{X_a}*{\cal{D}}\alpha),
\end{equation}
and
\begin{equation}
\hat\tau_{a}[dQ] =
\frac{1}{2} (\iota_{X_a}dQ \wedge *dQ - dQ \wedge \iota_{X_a}*dQ).
\end{equation}
The $\alpha$ field variation of  (\ref{act1}) yields
\begin{equation}    2 \alpha
R^{a}_{\,\,b} \wedge *(e_a \wedge e^b) + \c
{\cal{D}}*{\cal{D}}\alpha  = 0. \label{ALPHA}
 \end{equation}
The trace of (\ref{EIN1}) follows by  left exterior multiplication
by $e^a$. By comparing with  (\ref{ALPHA}) the terms proportional
to the scalar curvature can be eliminated yielding:
 \begin{equation}
\frac{\c}{2}{\cal{D}}*{\cal{D}}\alpha^2 = 0.\label{ALPHA1}
\end{equation}
Connection variations of  (\ref{act1}) are carried out under the
constraint (\ref{constraint}), a condition that may be imposed by
the method of Lagrange multipliers. Such variations need  care
since one cannot raise and lower indices freely under the
covariant derivatives. Noting that the variation $\delta Q =
-\frac{1}{4} \delta \Lambda^{a}_{\,\,b} \eta^{b}_{\,\,a}$  one
obtains
\begin{equation}\label{connvar}
2  D(\alpha^2 *(e_a \wedge e^b)) - \eta^{b}_{\,\,a}
( \c \alpha *{\cal{D}}\alpha + d*dQ ) = 0.
\end{equation}
 The symmetric part of (\ref{connvar})
gives the $Q$-field equation
\begin{equation}
 d*dQ + \c \alpha *{\cal{D}}\alpha =
0.\label{Q}
\end{equation}
Note that no further equations arise from exterior differentiation
of (\ref{Q}) since (\ref{ALPHA1}) is recovered. On the other hand,
by lowering an index in  (\ref{connvar}), picking up a new term
proportional to the non-metricity in the process, and setting the
anti-symmetric part of the resulting equation  to zero one finds:
\begin{equation}
d\alpha^2 \wedge *(e_a \wedge e_b) + \alpha^2
*(e_a \wedge e_b \wedge e_c) \wedge T^c + 2 \alpha^2 Q \wedge *(e_a
\wedge e_b) = 0.
\end{equation}
This is an algebraic equation that can be
solved uniquely for the torsion 2-forms: $$ T^a = e^a \wedge
\frac{d\alpha}{\alpha} + e^a \wedge Q.$$
The corresponding contortion 1-forms are
\begin{equation}
K_{ab} = - e_b\,\iota_{X_a}(\frac{d\alpha}{\alpha} + Q)
+ e_a\,\iota_{X_b}(\frac{d\alpha}{\alpha} + Q) .
\end{equation}
Therefore the
connection 1-forms become
\begin{equation}
\Lambda_{ab} = \omega_{ab} -
e_b\,\frac{\iota_{X_a}d\alpha}{\alpha} +
e_a\,\frac{\iota_{X_b}d\alpha}{\alpha} - \eta_{ab} Q.
\end{equation}
The curvature 2-forms $R^{a}_{\,\,b}$ of this connection may be
written in terms of the curvature 2-forms $R(\omega)^{a}_{\,\,b}$  of
the Levi-Civita  connection as follows:
\begin{eqnarray}
R^{a}_{\,\, b} &=& R(\omega)^{a}_{\,\, b} -
D(\omega)(\frac{\iota^a d\alpha}{\alpha}) \wedge e_b
+D(\omega)(\frac{\iota_b d\alpha}{\alpha}) \wedge e^a   \nonumber \\
&+& \frac{\iota^a d\alpha}{\alpha} \frac{d\alpha}{\alpha} \wedge e_b
- \frac{\iota_b d\alpha}{\alpha} \frac{d\alpha}{\alpha} \wedge e^a
\nonumber \\
&+& *( \frac{d\alpha}{\alpha} \wedge *\frac{d\alpha}{\alpha} ) e^a \wedge
e_b - \eta^{a}_{\,\, b} dQ . \label{ref1}
\end{eqnarray}
The {\sl (torsion-free) Weyl connection 1-forms} \cite{4a,4c,9a,10b} are determined by the
difference
\begin{equation}
\Gamma^{a}_{\,\,b} \equiv \Lambda^{a}_{\,\,b} - K^{a}_{\,\,b}.
\end{equation}
Thus
\begin{equation}
\Gamma_{ab} = \omega_{ab} +
e_b\,\iota_{X_a}Q -
e_a\,\iota_{X_b}Q - \eta_{ab} Q.
\end{equation}
The same  curvature 2-forms $R^{a}_{\,\,b}$ above may also be expressed in terms
of the curvature 2-forms $R(\Gamma)^{a}_{\,\,b}$  of this Weyl connection:
\begin{eqnarray}
R^{a}_{\,\, b} &=& R(\Gamma)^{a}_{\,\, b}
- D(\Gamma)(\iota^a (\frac{d\alpha}{\alpha}+ Q)) \wedge e_b
+ D(\Gamma)(\iota_b (\frac{d\alpha}{\alpha}+ Q)) \wedge e^a  \nonumber \\
&-& (\iota^a (\frac{d\alpha}{\alpha}+ Q)) (\frac{d\alpha}{\alpha}+ Q) \wedge e_b 
- (\iota_b (\frac{d\alpha}{\alpha}+ Q)) (\frac{d\alpha}{\alpha}+ Q) \wedge e^a \nonumber \\
&+& *((\frac{d\alpha}{\alpha}+Q) \wedge *(\frac{d\alpha}{\alpha}+Q)) e^a \wedge
e_b - 2 (\iota^a (\frac{d\alpha}{\alpha}+Q)) Q \wedge e_b .
\end{eqnarray}

The above field equations $(\ref{EIN1}),(\ref{ALPHA1}),(\ref{Q})$
constitute a Weyl covariant theory of gravity in which solutions
with a varying everywhere non-zero $\alpha$ can be used to
determine a dynamic units frame. A local Weyl gauge can be found
that transforms any such solution for $\alpha$  to a constant $
\alpha_0$. In such a gauge the equations above take the form:
\begin{equation}
{\alpha_0}^2 G_a = \c {\alpha_0}^2
\tau_a [Q] + \hat\tau_a [dQ] ,
\end{equation}
\begin{equation}
d*dQ + \c {\alpha_0}^2 *Q = 0 ,
\end{equation}
\begin{equation}
d*\,Q=0
\end{equation}
where
\begin{equation}
\tau_{a}[Q] = \frac{1}{2}( \iota_{X_a}Q * Q + Q \wedge
\iota_{X_a}*Q).
\end{equation}
It should be stressed that choosing such a gauge is in no way
equivalent  to breaking local Weyl symmetry.
It follows from $(\ref{ref1})$  that this is an Einstein-Proca system and
any solution to this  system will
generate a class of Weyl gauge equivalent solutions. Since only
the light-cone conformal spacetime structure is a Weyl class
invariant such a theory has no preferred mass scale.

One way to break the Weyl covariance of the above action is to
consider the theory with $Q=0$ and $\c\neq 0$. 
In this case it follows from (\ref{ref1}) that the gravitational field
equations can be written
in terms of the Einstein tensor of the Levi-Civita connection.
One then recovers
the Brans-Dicke theory (with the Brans-Dicke scalar $\alpha^2$) in
the absence of matter \cite{5a,5b,5c}. 
The Brans-Dicke coupling parameter $\omega$ is identified from
$\c = 2\omega +3$.
The effect of the contribution by
  the scalar function $\alpha$ in (\ref{ref1})
can be identified with the so-called improved
stress tensor \cite{8b,1a}.
We shall proceed differently and maintain
the presence of the geometrical field $Q$.

It is also instructive to express the  action (\ref{act1}) in terms of
the Weyl curvature scalar $R^{a}_{\,\,b}(\Gamma) \wedge *(e_a
\wedge e^b)$:
\begin{equation}
 \frac{\bfLam_1}   {\bfLam_0} = \frac{\alpha^2}{2}
R^{a}_{\,\,b}(\Gamma) \wedge *(e_a \wedge e^b) - \frac{\c-6}{2}
{\cal{D}}\alpha \wedge *{\cal{D}}\alpha 
 -\frac{1}{2} dQ \wedge *dQ  + mod(d).
\end{equation}
With $\c = 0$, this action was considered by Dirac in Ref.\cite{3b}
with an additional term proportional to $\alpha^4$
(considered in the next section).
It is also possible to express the action in terms of the
Levi-Civita curvature scalar  $R^{a}_{\,\,b}(\omega) \wedge *(e_a
\wedge e^b)$:
\begin{eqnarray}
\frac{\bfLam_1}   {\bfLam_0} &=& \frac{\alpha^2}{2}
R^{a}_{\,\,b}(\omega) \wedge *(e_a \wedge e^b) - \frac{\c-6}{2}
d\alpha \wedge *d\alpha - \frac{\c}{2} d\alpha^2 \wedge *Q
\nonumber \\
&-&  \frac{1}{2} dQ \wedge *dQ - \frac{\c}{2} \alpha^2 Q \wedge
*Q + mod(d).
\end{eqnarray}
With $Q = 0$ and $\c = 0$ this action was considered by Anderson in
Ref. \cite{2a}
as the scale invariant limit of the Brans-Dicke theory.

\section{Mass generation}

A theory with no explicit scale does not describe the world in its
current epoch. Particles are observed as field quanta with
definite masses and electric charges and the classical world is
distinguished from  quantum phenomena by actions that are large
compared with the Planck unit of action.
Inspired by the Abelian Higgs model in electrodynamics we now
enlarge  the theory to include a local U(1) symmetry group. Thus
 a real U(1) gauge connection 1-form  $A$ is introduced along with
 a complex scalar field $\beta$
 transforming as
\begin{equation}
\beta \mapsto \exp({-ie\theta}) \beta
\end{equation}
under
\def\maxfactor{{\frac{e_0^2}{\hbar^2\,c^2}}}
\begin{equation}
A \mapsto A + d\theta
\end{equation}
where $e$ is a dimensionless U(1) charge parameter. %
Naturally $[e\,\theta]=1 $ and the 1-form $A$ will be taken as
inert under Weyl scalings with $[A]=1$ while $\beta$ has constant
dimensionless Weyl charge $q$:
\begin{equation}
\beta \mapsto \exp({-q\sigma})
\beta.
\end{equation}
 Thus to maintain both these local symmetries  the full gauge 
covariant exterior  derivative of $\beta$ becomes:
\begin{equation}
{\cal{D}} \beta \equiv d\beta +ieA \beta + q Q \beta
\end{equation}
while its U(1) gauge covariant exterior derivative will be
written $$ D_A \beta = d \beta +ie A \beta. $$
With $\beta^\dagger \mapsto \exp({ie\theta}) \beta^\dagger$
 under $U(1)$ and
$$ {\cal{D}} \beta^\dagger \equiv d\beta^\dagger -ieA \beta +
q Q \beta^\dagger ,$$
 an additional Weyl and $U(1)$ invariant
contribution to the previous action with
 \begin{equation}\label{act2}
\bfLam_2 =\bfLam_0\left( - \frac{1}{2}
\alpha^{2-2q} {{\cal{D}} \beta}^{\dagger} \wedge *{\cal{D}} \beta
- \frac{1}{2}\,dA \wedge * dA - V(|\beta|,\alpha)
*1\right)
\end{equation}
 is now considered.  The Weyl charge
assignment for $\beta$ dictates $[\beta]=[\beta^\dagger]=L^{-q}.$
The Weyl and U(1) invariant interaction potential $V$, depending
on three dimensionless constants $\lamone, c_2, \lambda_3$, will
be responsible for the {\it simultaneous spontaneous breakdown} of
the U(1) and Weyl symmetries:
\begin{equation}
V(|\beta|,\alpha) = \lamone \alpha^{4-4q} {|\beta|}^4
\lamtwo \,\,\alpha^{4-2q} {|\beta|}^2 + \lambda_3 \alpha^4.
\end{equation}
One may envisage that the magnitudes and signs of the constants in $V$
control the breakdown as a function of epoch. In the context of
cosmology their values might be determined by ``matter'' or
``radiation'' temperature. Thus we expect stationary field
configurations to occur for different minima of such a potential.
By linearising the theory about such solutions we effectively
break the above symmetries and seek a mass spectrum for the scalar
and gauge field excitations in such a background. The role of the
$\alpha$ field in maintaining the Weyl invariance of $V *\,1$ in
(\ref{act2}) should be noted in this context.

The field equations  follow from varying the action
\begin{equation}
\S[\{e^a\}, \{\Lambda^a{}_b\},\alpha,\beta,A]=\int_M
 (\bfLam_1+\bfLam_2)
\end{equation}
 under the same connection constraint as before.
The gravitational  field equations  are
\begin{equation}
\alpha^2 G_a =
\c \tau_{a}[\alpha] + \hat\tau_{a}[dQ] + \alpha^{2-2q}
\tau_{a}[\beta] + \hat\tau_{a}[dA] - V  *e_a.
\end{equation}
where now
$$ \tau_{a}[\beta] = \frac{1}{4}
( \iota_{X_a}{{\cal{D}}\beta}^{\dagger} *{{\cal{D}}\beta} +
{{\cal{D}}\beta}^{\dagger}
\wedge \iota_{X_a}*{{\cal{D}}\beta}
+ \iota_{X_a}{{\cal{D}}\beta} *{{\cal{D}}\beta}^{\dagger} +
{{\cal{D}}\beta}
\wedge \iota_{X_a}*{{\cal{D}}\beta}^{\dagger}).$$
The $\alpha$-field equation is
\begin{equation}
{\frac{\gamma}{2}}{\cal{D}}*{\cal{D}}\alpha^2 + q \alpha^{2-2q}
({{\cal{D}} \beta})^{\dagger} \wedge \,*\,{\cal{D}} \beta + \left
( 4 V - \alpha \frac{\partial V}{\partial \alpha} \right ) *1 =0.
\end{equation}
The connection field equations  include
\begin{equation}
d*dQ + \gamma\alpha *{\cal{D}}\alpha + \frac{q}{2}
\alpha^{2-2q} *{\cal{D}}(|\beta|^2) = 0
\end{equation}
together with the torsion 2-forms
$$ T^a = e^a \wedge \frac{d\alpha}{\alpha} + e^a \wedge Q.$$
The $\beta$  field equation  obtained by varying
$\beta^{\dagger}$ is:
\begin{equation}
\frac{1}{2} {\cal{D}}(\alpha^{2-2q} *{\cal{D}} \beta) =
\frac{\partial V}{\partial \beta^{\dagger}} *1
\end{equation}
 and the $A$ field
equation is
\begin{equation}
d*dA + i\frac{e}{2} \alpha^{2-2q}
*\left (({D_{A} \beta})^{\dagger} \beta -\beta^{\dagger} (D_A
\beta) \right ) = 0.
\end{equation}

If one  writes
\begin{equation}
\beta = |\beta| e^{-ie \phi}
\end{equation}
with $[e\phi]=1$
so that under local scalings $ |\beta| \mapsto e^{-q\sigma}
|\beta| , \phi \mapsto \phi$, then under a local U(1) phase change
$|\beta| \mapsto |\beta|  , \phi \mapsto \phi + \theta.$ Note that
the combination $B \equiv A - d\phi$  appearing in all field
equations is invariant under  both scalings and phase
transformations and  the $A$-field equation now reads
\begin{equation}
 d*dB + e^2 \alpha^{2-2q} {|\beta|}^2 *B = 0,
\end{equation}
 while the $\beta$ field equation reduces to
\begin{equation}
{\cal{D}}(\alpha^{2-2q} *{\cal{D}}{|\beta|}) - e^2 |\beta|
\alpha^{2-2q} B \wedge *B = \alpha^{4-4q} |\beta| \left ( 2
\lamone {|\beta|}^2  \lamtwo \alpha^{2q} \right ) *1.
\end{equation}

We are interested in an epoch where  $\lamone > 0 , c_2 > 0$.
A  stationary ground state solution arises with 
${\cal{D}}\alpha = 0$ and $ {\cal{D}}\beta = 0$.
Since ${\cal{D}}^{2} \alpha = dQ \alpha$,  these solutions
 for $\alpha \neq 0$ will satisfy  $dQ =0$.
Thus we represent this solution in a gauge with  
$\alpha = \alpha_0$,   $Q = 0$ , $B = 0$ and   
$|\beta| = |\beta_0|$   (with the phase of
$\beta$ arbitrary)  such that
\begin{equation}
{|\beta_0|}^2
= \frac{c_2}{2 \lamone} {\alpha_0}^{2q}
\end{equation}
with $$\lambda_3
=\frac{{c_2}^2}{4 \lamone}.$$ The latter choice selects a
Minkowski metric solution $g = \eta$ corresponding to a
Levi-Civita flat ground state solution \footnote{By modifying $V$
one might consider symmetry breaking about a de Sitter
background.}. 
Thus the effective  potential takes the form
\begin{equation}
V(|\beta|,\alpha) = \lamone \alpha^4 \left ( \alpha^{-2q}
{|\beta|}^2  - \frac{c_2}{2 \lamone} \right )^2.
\end{equation}
To determine the mass spectrum in the broken symmetry phase the
field equations  
must be appropriately linearised about the ground state solution.
This is achieved  by writing
\begin{eqnarray}
\alpha = \alpha_0 + \epsilon \hat{\alpha} \, \, , \, \, 
Q &=& \epsilon \hat{Q} \, \, ,
\, \,  g = \eta + \epsilon^2 \hat{g}, \nonumber \\
 B &=& \epsilon \hat{B}, \nonumber \\
\beta = ( |\beta_0|
&-& \lambda \epsilon \hat{\alpha} ) \exp(-ie\phi).
\end{eqnarray}
where $\lambda =
\frac{\gamma}{q} {\alpha_0}^{q-1} \sqrt{\frac{2 c_1}{c_2}}$. 

Then to order $\epsilon$ the wave equation for  $\hat{Q}$ is
\begin{equation}
\mbox {{\tt *}}d\mbox {{\tt *}}d\,\hat{Q}   +
( \gamma\,{\alpha_0}^2 + \frac
{q^2 {\alpha_0}^2 c_2}{2 c_1}) \hat{Q}  =  0. \label{81}
\end{equation}
This implies
\begin{equation}
d*\hat{Q} = 0.
\end{equation}
Eqn.(\ref{81})  admits propagating modes with angular frequency $\omega$
and wave number $k$ satisfying the dispersion relation:
\begin{equation}
{\omega}^{2}={c}^{2}{k}^{2}+{c}^{2}\gamma\,{a_{{0}}}^{2}
+{\frac {{a_{{0}}}^{2}{c}^{2}{q}^{2}c_{{2}}}{2 c_{{1}}}} .
\end{equation}
Similarly to order $\epsilon$ the wave equation for  $\hat{B}$ is
\begin{equation}
\mbox {{\tt *}} d
\mbox {{\tt *}}  d \hat{B }   = -{\frac {{{\it
e}}^{2}{\alpha_{{0}}}^{2}c_{{2}}}{2 c_{{1}}}} \hat{B}
\end{equation}
with dispersion relation
\begin{equation}
{\omega}^{2}={c}^{2}{k}^{2} + {\frac {{{\it
e}}^{2}{c}^{2}{\alpha_{{0} }}^{2}c_{{2}}}{2 c_{{1}}}} .
\end{equation}
Thus in a units frame with constant $\hbar$ one  predicts for
$\gamma > 0$,  vector particles having  positive gravitational
masses given by:
\begin{equation}
M_{\hat{Q}}^2=\frac{\alpha_0^2\hbar^2}{c^2}\left(\gamma+\frac{c_2q^2}{2c_1}
\right) ,
\end{equation}
\begin{equation}
M_{\hat{B}}^2=\frac{\hbar^2 e^2 \alpha_0^2 c_2}{2c^2 c_1} .
\end{equation}

\def\ha{{\hat{\alpha}}}
\def\hQ{{\hat{Q}}}

\noindent Furthermore using (\ref{81})  equations 
(71), (77)  for   $\alpha$ and $\beta$  both reduce  
to order $\epsilon$ to
\begin{equation}
*d*d\,\ha  + ( 4 c_2 {\alpha_0}^2 +
\frac{2\alpha_0^2 q^2 c_2^2}{c_1\gamma} ) \,\ha = 0. 
\end{equation}
with dispersion relation
\begin{equation}
{\omega}^{2}={c}^{2}{k}^{2}+2\,{\frac {{\alpha_{{0}}}^{2}{c_{{2}}}^
{2}{c}^{2}{q}^{2}}{\gamma\,c_{{1}}}}+4\,{\alpha_{{0}}}^{2}c_{{2}}{c}^{2}.
\end{equation}
Thus the spontaneous breakdown induces a massive excitation with
\begin{equation}
M^2_{\ha}=\frac{4c_2\alpha_0^2\hbar^2}{c^2}\left(1+\frac{c_2q^2}{2\gamma
c_1} \right)
\end{equation}
for the $\ha$ field.
We note that the masses in the broken phase are determined by the
parameters $\alpha_0, \gamma, c_1, c_2, e, q$. The scale of the broken 
theory can be established in terms of any one of these masses.

\section{Mass and Charge Generation}

In the previous section the $U(1)$ symmetry was broken in the
process of mass generation. Consequently one cannot identify
electric charge in the broken phase. The ``standard model''
however provides a symmetry breaking mechanism that leaves a
$U(1)$ symmetry intact compatible with the observation of electric
charge conservation. It is of interest to embed this mechanism
into a theory with local Weyl symmetry. We restrict to the bosonic
sector of the $SU(2)_{I} \times U(1)_{Y}$ electroweak  theory
which contributes an action  4-form
\begin{eqnarray}
\frac{\bfLam_3}{\bfLam_0} &=&
- \frac{1}{2} dA \wedge *dA - \frac{1}{2} Tr({\bf F}\wedge *{\bf
F}) \\
&-& \frac{\alpha^{2-2q}}{2} ({\cal{D}}{\bf \Phi})^{\dagger}
\wedge *({\cal{D}}{\bf \Phi}) - V(|{\bf \Phi}|,\alpha) *1 , \nonumber
\end{eqnarray}
where $iA$ is the hypercharge potential 1-form, ${\bf F} = d{\bf
A} + [{\bf A}, {\bf A}]$ with ${\bf A}$ being the $SU(2)$ Lie
algebra (with basis ${\bf T}_j$) valued potential 1-form. Here the
Higgs scalar ${\bf \Phi}$ is a complex isodoublet
\begin{equation}
{\bf \Phi} =
\left (
\begin{array}{c} \phi_{+} \\ \phi_0 \end{array} \right )
\end{equation}
carrying a Weyl representation  with Weyl charge $q$
\begin{equation}
{\bf \Phi}
\mapsto e^{-q\sigma} {\bf \Phi},
\end{equation}
 which under a $SU(2) \times
U(1)$ transformation transforms as
\begin{equation}
{\bf \Phi} \mapsto e^{ -g {\bf \Theta} \cdot {\bf T} -
\frac{i}{2}g^{\prime} \theta } {\bf \Phi} .
\end{equation}
Thus its gauge  covariant exterior derivative
\begin{equation}
{\cal{D}} {\bf
\Phi} = d{\bf \Phi} + g {\bf A} {\bf \Phi} + i\frac{g^{\prime}}{2}
A {\bf \Phi} + q Q {\bf \Phi}
\end{equation}
where $${\bf A} = A_{j} \frac{{\bf
t}_j}{2i}$$ with  Pauli matrices  $ \{ {\bf t}_j\} $. To maintain
Weyl invariance  the  potential is constructed as
\begin{equation}
V(|{\bf \Phi}|,\alpha) = \lambda \alpha^4 ( \alpha^{-2q}{|{\bf \Phi}|}^2 -
v^2 )^2
\end{equation}
where $\lambda$ and $v$ are real constants. The
variational field equations are found from the combined action
 \begin{equation}\label{act3}
\S[\{e^a\}, \{\Lambda^a{}_b\},\alpha,{\bf \Phi},{\bf A},A]=\int_M
 (\bfLam_1+\bfLam_3) .
\end{equation}
Coframe variations yield the gravitational field equations
\begin{equation}
\alpha^2 G_a = \gamma \tau_{a}[\alpha] + \hat\tau_{a}[dQ] +
\tau_{a}[{\bf \Phi}] + \hat\tau_{a}[dA] + \hat\tau_{a}[{\bf F}] -
V(|{\bf \Phi}|,\alpha) *e_a ,
\end{equation}
while  $\alpha$ variations give:
\begin{equation}
\frac{\gamma}{2} {\cal{D}}*{\cal{D}}\alpha^2 + q \alpha^{2-2q}
({\cal{D}} {\bf \Phi})^{\dagger} \wedge *{\cal{D}} {\bf \Phi} + (
4V - \alpha \frac{\partial V}{\partial \alpha} )*1 = 0.
\end{equation}
The connection variational equation $$ D(\alpha^2 *(e_a \wedge
e^b)) - \eta^{b}_{\,\,a} \left [ \gamma \alpha
*{\cal{D}}\alpha + d*dQ + q \alpha^{2-2q}  *\left ( {\bf
\Phi}^{\dagger} {\cal{D}} {\bf \Phi} + ({\cal{D}} {\bf
\Phi})^{\dagger} {\bf \Phi} \right ) \right ] = 0 ,$$ can be
decomposed into its symmetrical and anti-symmetrical parts as
before. The antisymmetrical part is solved for the torsion 2-forms
as $$ T^a = e^a \wedge \frac{d\alpha}{\alpha} + e^a \wedge Q,$$
while the symmetrical part gives the $Q$-field equation
\begin{equation}\label{Qeqn}
d*dQ + \gamma \alpha *{\cal{D}}\alpha + q
\alpha^{2-2q}  *\left ( {\bf \Phi}^{\dagger} {\cal{D}} {\bf \Phi}
+ ({\cal{D}} {\bf \Phi})^{\dagger} {\bf \Phi} \right ) = 0.
\end{equation}
 We note that the
potentials $A,{\bf A}$ decouple from the $Q$-field equation
(\ref{Qeqn}). These gravitational field equations are coupled to
the Yang-Mills-Higgs  equations  obtained by varying ${\bf
\Phi},{\bf A},A$ :
 \begin{equation}\label{YMeqn}
{\cal{D}}_{A} *{\bf F} + \frac{g}{2}
\alpha^{2-2q} *\left ( ({\cal{D}} {\bf \Phi})^{\dagger}\frac{{\bf
t}_j}{2i} {\bf \Phi} - {\bf \Phi}^{\dagger} \frac{{\bf t}_j}{2i}
{\cal{D}} {\bf \Phi} \right ) {\bf T}_j = 0 ,
\end{equation}
\begin{equation}\label{Aeqn}
d*dA -i\frac{g^\prime}{4} \alpha^{2-2q} *\left ( ({\cal{D}} {\bf
\Phi})^{\dagger} {\bf \Phi} - {\bf \Phi}^{\dagger} {\cal{D}} {\bf
\Phi} \right ) = 0 ,
\end{equation}
\begin{equation}\label{Phieqn}
\frac{1}{2} {\cal{D}}(\alpha^{2-2q} *{\cal{D}}  {\bf \Phi}) -
\frac{\partial V}{\partial {\bf \Phi}^{\dagger}}*1 = 0.
\end{equation}
By contrast $Q$  decouples from both the gauge  field equations (\ref{YMeqn}), (\ref{Aeqn}).

A stationary vacuum solution is $g = \eta, {\bf A} = 0, A = 0,
 Q = 0, \alpha = \alpha_0$ and
\begin{equation}
{\bf \Phi}_0 = \left ( \begin{array}{c} 0\\v \end{array}
\right ) {\alpha_0}^q ,
\end{equation}
so that $|{\bf \Phi}_0|^2 = v^2
{\alpha_{0}}^{2q}.$ 
As in the previous section, to break the symmetry group an effective
linearisation about this vacuum is required:
\begin{eqnarray}
\alpha = \alpha_0 + \epsilon
\hat{\alpha} \quad , \quad
  Q &=& \epsilon \hat{Q} \quad , \quad
 g = \eta + \epsilon^2 \hat{g} , \nonumber \\
{\bf A} = \epsilon \hat{{\bf A}} \quad &,&
\quad A = \epsilon \hat{A} ,  \\
{\bf \Phi} &=& \left ( \begin{array}{c}
\epsilon {\hat{\phi}}_{+} \\
v {\alpha_{0}}^{q} - \frac{\gamma {\alpha_0}^{q-1}}{q v}  \epsilon \hat{\alpha} \end{array} \right )
\nonumber
\end{eqnarray}
where $\hat{\alpha}$ is real and ${\hat{\phi}}_{+}$ is complex.
The linearised $Q$-field equation is
\begin{equation}
*d*d\hat{Q} + (\gamma {\alpha_0}^2 + q^2 {\alpha_0}^2 v^2 ) \hat{Q} = 0
\end{equation}
implying $d*\hat{Q} = 0$,  
and the Proca mass for the $\hat{Q}$ field  is given exactly as in the
last section:
\begin{equation}
{M_{\hat{Q}}}^2 = \frac{\hbar^2 {\alpha_0}^2}{c^2} ( \gamma + q^2 v^2).
\end{equation}
The linearisation of the $A$ and ${\bf A}$ field equations follows
as in standard electro-weak theory yielding $W$-bosons  $W^{\pm} =
\hat{A}_1 \pm i \hat{A}_2$ with masses
\begin{equation}
{M_W}^2 = \frac{\hbar^2 v^2}{4c^2}
{\alpha_0}^2 g^2,
\end{equation}
a $Z$-boson $ Z^0 = g \hat{A}_3 -g^{\prime} \hat{A}$ with  mass
\begin{equation}
{M_Z}^2 = \frac{\hbar^2 v^2}{4 c^2} {\alpha_0}^2 (g^2 +
{g^{\prime}}^2),
\end{equation}
and a massless photon $\gamma = g^{\prime} \hat{A}_3 + g \hat{A}$,
\begin{equation}
M_{\gamma} = 0.
\end{equation}
The $W^{\pm}$ and $Z^0$ fields satisfy 
$d*W^{\pm} = 0$ and $d*Z^0 = 0$.
As a result of the residual $U(1)$
local symmetry the massless photon gives rise to electric current
conservation and the identification of electric charge.

It follows from above that the linearised field equations for 
${\hat{\phi}}_{+}$ and $\hat{\alpha}$  are decoupled:
\begin{equation}
*d*d\hat{\alpha} + ( \frac{ 8q^2 \lambda {\alpha_0}^2 v^4}{\gamma}
+  8 \lambda {\alpha_0}^{2} v^2) \hat{\alpha} = 0 ,
\end{equation}
\begin{equation}
*d*d{\hat{\phi}}_{+} = \frac{vg{\alpha_0}^{q}}{2} *d*(\hat{A_1}
 -i\hat{A_2}) \label{4.9} .
\end{equation}
From $(\ref{4.9})$ we note  that the ${\hat{\phi}}_{+}$-excitation
is not independent and can be determined in terms of the
$SU(2)$-potentials and appropriate boundary conditions.  The
remaining equation for  $\hat{\alpha}$ determines the 
scalar boson mass
in terms of the parameters of this model:
\begin{equation}
{M_\ha}^2 = \frac{\hbar^2 8 \lambda {\alpha_0}^2 v^2}{c^2 } (
1 + \frac{q^2 v^2}{\gamma}).
\end{equation}
In this picture the mass of the
Higgs scalar $\ha$ depends on both the Weyl charge $q$ of
${\bf\Phi}$ and the ``dilaton'' coupling $\gamma$, an explicit
recognition of its gravitational pedigree.

\section{Conclusion}

It has been shown that the breakdown of local Weyl symmetry in a
theory of gravity can be accommodated in the context of the
standard model of particle interactions. A natural setting for
this mechanism is a spacetime geometry described by a connection
with dynamical torsion and a metric that is not covariantly
constant. Together with a scalar field such a connection encodes
new gravitational interactions that can be reformulated in terms
of the standard description of Einsteinian gravity. The emergence
of spacetime torsion, dependent on the gradient of the dynamic
scalar field, is responsible for the appearance of the so-called
{\sl improved stress-energy tensor}. It has long ago been noted
\cite{8b} that this consequence of Weyl symmetry results in
``improvements'' to perturbative calculations involving gravitons.
In the broken
 phase in which electroweak phenomenology is discussed the theory
 gives rise to a Higgs particle with mass $M_\eta$ and a new
 electrically neutral vector boson with mass $M_Q$ such that
 $$\frac{M_\ha^2}{M_Q^2}=\frac{8\lambda v^2}{\gamma}$$
 in terms of the couplings in the theory.
It is of interest to note that a number of grand unified models
predict a new neutral vector boson and according to \cite{15b}
experimental data are now detailed enough to check for its
existence. It appears that such data are better described if the
presence of such a boson is assumed.

The theory in this paper has been analysed in a broken phase in
which normal gravitation (based on metric perturbations about
Minkowski spacetime) is negligible. The mass generation mechanism
has been connected  with a component of non-Einsteinian
gravitation associated with the Weyl 1-form $Q$. 
Although the relevance of  Weyl symmetry to mass
generation has been noted before \cite{13a,13b} we believe that the
approach adopted in this paper is new.
The Weyl 1-form  is part of
the natural spacetime geometry determined from our action
principle and may be expected to give rise to new kinds of force
on classical particles or in cosmological dynamics. The
interaction potential $V$ that simulates the symmetry breakdown is
also dependent on a ``dilatonic'' scalar and this can play an
intimate role in the non-perturbative aspects of the theory. As
has been noted elsewhere such scalars may determine the dependence
of certain ``constants'' of nature on the cosmological epoch or
other gravitational phases. However, if the neutral boson described
by the excitation of the Weyl potential $Q$ in Minkowski spacetime
can be observed in current electroweak data it may signal that a
new component of gravitation can influence phenomenology at
energies well above the Planck scale.

\bigskip 

\section{Acknowledgements}
Both authors are grateful to the Leverhulme Trust and RWT is
grateful to BAe-Systems  for support for this research.

\newpage
{\small
\begin{thebibliography}{99}

\bibitem{4a}  H. Weyl, Preuss. Akad. Wiss., Sitzungsber. (1918) 465

\bibitem{4b} H. Weyl, Z. Phys. {\bf 56}(1929)330

\bibitem{4c} A. S. Eddington,{\bf The Mathematical Theory of Relativity}
\\(Cambridge U. P., 1923)

\bibitem{4d}  E. Cartan, Ann. Sci. Ec. Norm. Sup. {\bf 60}(1943)1

\bibitem{book} I. M. Benn, R. W. Tucker, {\bf An Introduction to
Spinors and Geometry with Applications in Physics} (Adam Hilger, 1987)

\bibitem{16c} E. J. Post, {\bf Formal Structure of Electromagnetics} 
\\(Dover, 2nd Edition, 1997)

\bibitem{16d} P. Szekeres, Int. J. Theo. Phys. {\bf 17}(1977)957

\bibitem{3a}  P. A. M. Dirac, Proc. Roy. Soc. {\bf A165}(1938)199

\bibitem{3b} P. A. M. Dirac, Proc. Roy. Soc. {\bf A333}(1973)403

\bibitem{3c} P. A. M. Dirac, Proc. Roy. Soc. {\bf A365}(1979)19

\bibitem{2c} L. L. Smalley, Phys. Rev. {\bf D33}(1986)3590

\bibitem{9a} K. P. Tod, J. London Math. Soc. {\bf 45}(1992)341

\bibitem{10b} K. P. Tod, Class. Q. Grav. {\bf 13}(1996)2609

\bibitem{5a} C. Brans, R.H. Dicke, Phys. Rev. {\bf 124}(1961)925

\bibitem{5b} R. H. Dicke, Phys. Rev. {\bf 125} (1961) 2163

\bibitem{5c} R. H. Dicke in
{\bf Relativity, Groups and Topology}\\ 1963 Les Houches Lectures,
Edited by B. S. De Witt (North-Holland, 1964)

\bibitem{8b} S. Coleman, R. Jackiw, Ann. Phys. {\bf 67}(1971)552

\bibitem{1a} T. Dereli, R. W. Tucker, Phys. Lett. {\bf B110}(1982)133

\bibitem{2a} J. L. Anderson, Phys. Rev. {\bf D3}(1971)1689

\bibitem{15b}J.Erler, P. Langacker, Phys. Rev. Lett. {\bf 84}(2000)212

\bibitem{13a} W. Drechsler, H.Tann, Found. Phys. {\bf 29}(1999)1023
 (gr-qc/9802044)

\bibitem{13b} W. Drechsler, Found.Phys. {\bf 29}(1999)1327  (gr-qc/9901030)

\end{thebibliography} }
\end{document}